# Towards Optimal Broadcast in Wireless Networks

Zygmunt J. Haas, *Fellow*, *IEEE*, and Milen Nikolov

*Abstract*— Broadcast is a fundamental operation in networks, especially in wireless Mobile Ad Hoc NETworks (MANET). For example, some form of broadcasting is used by all on-demand MANET routing protocols, when there is uncertainty as to the location of the destination node, or for service discovery. Being such a basic operation of the networking protocols, the importance of efficient broadcasting has long been recognized by the networking community. Numerous papers proposed increasingly more efficient implementation of broadcasting, while other studies presented bounds on broadcast performance. In this work, we present a new approach to efficient broadcast in networks with dynamic topologies, such as MANET, and we introduce a new broadcasting algorithm for such networking environments. We evaluate our algorithm, showing that its performance comes remarkably close to the corresponding theoretical performance bounds, even in the presence of packet loss due to, for example, MAC-layer collisions. Furthermore, we compare the performance of the proposed algorithm with other recently proposed schemes, including in various mobility settings.

## I. INTRODUCTION

BROADCAST is a fundamental network operation allowing a source node to send a message to all other nodes in the network. In the context of *Mobile Ad Hoc Networks* (*MANET*), where topology can change rapidly, where all communications are carried over a wireless medium, and where the network nodes are limited in energy and computational power, an efficient broadcast mechanism is exceptionally important for the overall network performance.

For instance, among the various proposed MANET routing protocols (e.g., AODV, DSR, OLSR, TRBF, ZRP), a prominent sub-group, referred to as *on-demand* or *reactive* routing protocols, is designed based on the philosophy that the discovery of a route in the network should be done only when there is an actual need to route traffic. The route discovery mechanism in on-demand routing protocols relies on some variant of broadcasting to locate a path between the source and the destination nodes. As another example, in highly reconfigurable topologies, where the lifetime of network routes may be shorter than the duration of a communication (especially in the case of connection-oriented communication) broadcast, by itself, could be used as a routing mechanism. Yet in other scenarios, when data dissemination to all the nodes in a MANET is needed, broadcast is an obvious solution.

The required features of an efficient broadcast algorithm in the context of wireless reconfigurable networks are that it:

(1) reaches *all* the network nodes;

(2) transmits the message as few times as possible (or, equivalently, reduces the number of times that the broadcast message is received by a network node, optimally to only once);

(3) minimizes delay (i.e., the time needed for the broadcast message to be received by the entire network);

(4) requires only locally available information (e.g., only the knowledge of the 1-hop topology);

(5) minimizes the effects on (1), (2), and (3) above due to topological changes during a broadcast propagation because of mobility, and due to packet loss (e.g., due to MAC-layer collisions).

Being such an essential network operation, it is not surprising that the importance of an efficient broadcast implementation has been widely accepted by the networking community. Unfortunately, few of the so-far-proposed approaches to the broadcast problem satisfy all the above requirements. Flooding-based protocols violate (2) above; furthermore, they lead to the notorious "broadcast storm" problem [1]. Probabilistic schemes, often based on percolation [3 − 9] do not satisfy (1) above.[1] Efficient backbone-based algorithms have also been proposed (e.g., the reader is referred to [46] for comparison of such schemes). They rely on finding minimum connected dominating sets (MCDS), which are constructed by identifying dominating sets, maximal independent sets, or Steiner trees *prior* to broadcast. While satisfying (1) − (3) above, such algorithms do not tolerate dynamic topologies well [19 − 22], thus are incompliant with (5). Furthermore, some of these algorithms are centralized [29, 30], thus violating (4). One recent example of a broadcast algorithm that satisfies (1) − (5) has been described in [25], though the algorithm requires the knowledge of the nodes' geographical position. (Section VIII provides more details and references to related work in the technical literature.)

The main contributions of this paper include the design of a novel broadcasting algorithm that, while utilizing only 1-hop local information, *simultaneously* achieves full coverage of the network with close to optimal number of broadcast messages

This work was supported by the NSF grant numbers ANI-0329905 and CNS-0626751, and by the AFOSR contract number FA9550-09-1-0121/Z806001.

Z. J. Haas is with the School of Electrical and Computer Engineering, Cornell University, Ithaca, NY 14853 USA (e-mail: haas@ece.cornell.edu).

M. Nikolov, is with the School of Electrical and Computer Engineering Cornell University, Ithaca, NY 14853 USA (e-mail: mvn22@ cornell.edu).

[1] To ensure full coverage, a stochastic scheme would need to transmit with probability close to 1, in which case the scheme degrades to flooding.

and with low delay. Furthermore, the algorithm is robust to rapid topological changes and to severe packet loss (*c.f.* (1) − (5) above). Moreover, the algorithm is simple to implement and does not require positional information, which in many network settings is infeasible or costly to obtain. In summary, the proposed algorithm is efficient, distributed, and relies only on local coverage information. To the best of our knowledge, previously suggested schemes have *not* satisfied *all* of these requirements of practical broadcast.

The algorithm is based on a distributed greedy heuristic to approximate the execution of the centralized greedy algorithm, thus allowing to practically tackle the NP-hard [32] problem of finding the *Minimum Connected Dominating Set* (MCDS).

We model, simulate, evaluate and compare the proposed broadcast scheme with the most efficient schemes to-date in the technical literature, and demonstrate that the proposed scheme outperforms them in the metrics considered.

Section II describes the system model and related assumptions. Section III explains the basic intuition behind the scheme. More details and an execution example of the scheme are provided in Section IV. Section V comparatively studies the performance of the scheme in various networking scenarios, including realistic mobility models (e.g., both the *Gauss-Markov* [33, 34] and the *Reference Point Group* [35] mobility models are considered), as well as in the presence of packet loss (e.g., due to MAC-layer collisions). Sections VI and VII conclude the paper, placing it in the context of future and prior work, respectively. In the Appendix, closed-form bounds on wireless broadcast efficiency are derived, providing benchmarks for broadcast algorithms.

## II. The System Model

The network model consists of $N$ equal-capability nodes with unique IDs, randomly distributed in a 2D plane.[2] The transmission range of all nodes is $r$ [meters]. Two nodes are referred to as 1-*hop neighbors* (or simply as *neighbors*) and can communicate directly if the distance between them is less than $r$ [meters]. Thus, the network is modeled as a Unit Disk Graph (UDG).

On the MAC layer, links are bidirectional and nodes share a single wireless channel. We consider two scenarios here: a) a *perfect* MAC layer to isolate other effects (e.g., collisions), so that the broadcast performance metrics reflect only the algorithmic efficiency; and b) *packet loss* (at the MAC and lower layers) to evaluate the performance of the algorithm in practical network settings, where collisions are present.

The system operation is time-slotted, and the network nodes are assumed to be only coarse-grain synchronized. The latter is a standard assumption of many distributed algorithms and can be implemented in variety of ways. For instance, distributed, control-message-based coarse-grain synchronization in multi-hop wireless networks would suffice; such schemes have been studied extensively in the literature (e.g., [28]). Recent advances in radio technologies could also be utilized (e.g., [27]). Finally, we assume that all nodes are cooperative and trust each other.

We represent the network as a connected graph $G=(S, E)$,

where $S$ is the set of all the network nodes and $E$ is the set of all the links and a link connects two nodes in $S$. Also, we label the broadcast source node as $s_0$. Next, we provide a few definitions, which will facilitate the description of the proposed scheme.

*Definition 1:* A *broadcast session* is the operation (including all related events) of delivering a message, created at one node – the *source* – to all the other network nodes.

*Definition 2:* A *covered node* is a node that has already received the broadcast message in a prior transmission of the broadcast session.

The source node of a broadcast session is always covered. A node that has not received the broadcast message at a particular time is referred to as an *uncovered* node at that time.

*Definition 3:* The *residual coverage* (*RC*) of a covered node $s$ ($s \in S$) at a particular time, referred to as $RC(s)$, equals the number of its 1-hop uncovered neighbors at that time.

*Definition 4:* We define $C$ as the set of all covered nodes at a particular time; and $Q$ as the set of nodes that have already transmitted the message at a particular time. We further define $NE(s)$ as the set of all the neighbors of the node $s$ ($s \in S$). We note that at any time, $Q \subseteq C \subseteq S$ and that $|S| = N$.

## III. The Solution Demystified

The problem of finding the most efficient broadcast scheme is equivalent to finding an approximation of the MCDS, while satisfying (1) − (5). And since finding the MCDS is an NP-hard problem, one needs to consider an appropriate heuristic, with the centralized *greedy algorithm* being one such an alternative (e.g., [31]). The basic idea of the greedy algorithm to find the MCDS is to repeatedly select nodes for transmission, such that in each round a node whose transmission *covers* the largest number of uncovered nodes is selected. Thus, each transmission "removes" the largest possible number of nodes from the set of uncovered nodes and, eventually, results in covering the whole network with a minimum number of transmissions. However, such a centralized greedy algorithm violates the requirement (4) of Section I. Consequently, we propose in this paper a particular distributed heuristic which approximates the operation of the centralized greedy algorithm. The operation of this distributed greedy heuristics relies on each node "scheduling" its transmission based on the value of its $RC$ – the larger is the value of $RC$, the sooner the node is scheduled to transmit.

To explain the operation of the proposed distributed heuristic, we consider first the operation of a centralized greedy scheme. We start with the initial set of covered node $C=\{s_0\}$ and $Q=\varnothing$. The source node transmits first, covering its neighbors: $C = \{s_0 \cup NE(s_0)\}$, $Q=\{s_0\}$. An "oracle" greedily chooses a node, $s_1$, from the set $C$-$Q$ with the largest $RC$ value to broadcast next; i.e., $\forall s \in (C$-$Q)$, $RC(s) \leq RC(s_1)$. After $s_1$ transmits, $C \leftarrow C \cup NE(s_1)$ and $Q \leftarrow Q \cup \{s_1\}$. Then, repeatedly, the next node with the largest $RC$ is selected to transmit from the set $C$-$Q$, until all the network nodes are covered; i.e., until $C=S$, at which time the algorithm terminates. The total number of transmissions in a broadcast session equals $|Q|$ at the algorithm termination time. Furthermore, the choice of node $s_i$ to transmit in the $i^{th}$ iteration allows maximizing the number of covered nodes during the $i^{th}$ transmission. This intuitively tends to minimize the total number of transmissions during the operation of the algorithm. Of course, the algorithm does not guarantee such a minimum, as in some cases choosing a node with smaller *RC* value first could, in fact, result in finding nodes with much larger *RC* value later, so that the overall number of transmissions is smaller.

---

[2] However, the results in this paper apply to 1D and to 3D networks, as well.

Although in [31], the authors discuss the inefficiency of a greedy scheme in finding MCDS. the above centralized greedy heuristic finds on the average a rather close approximation of a MCDS[3] in UDG as demonstrated in Section V. Of course, the challenge, similarly to other efficient MCDS approximation schemes, is to implement the "oracle" in a distributed manner; i.e., ordering nodes' transmissions based on their *RC* values, while utilizing only local topological information.

The following *Time Sequence Scheme* approximates the centralized greedy transmissions' order in time, by allowing nodes with larger *RC* values to transmit before nodes with smaller *RC* values. The timing of transmissions is enforced by a particularly structured sequence of time-slots, which is locally available to each node, and based on which each node is able to appropriately schedule its own transmission. The scheduling procedure is based on associating a particular *RC* threshold with each time-slot. Only nodes with *RC* values not smaller than this time-slot's *RC* threshold are allowed to transmit in the current time-slot.

## IV. THE TIME-SEQUENCE SCHEME

Consider the following scheme which attempts to order the transmission of the nodes, so that nodes with larger *RC* transmit first. Assume that each subsequent time-slot is associated with a smaller *threshold* value of *RC,* and also that in a particular time-slot only nodes with *RC* values of at least as large as the time-slot's threshold are allowed to transmit. Upon receiving a broadcast message, a node marks itself as covered, determines its current *RC,* and schedules itself to transmit in a future time-slot, depending on its current *RC*. However, such a naïve implementation does not take into consideration the fact that, with each transmission, the set of newly covered nodes may have *RC* values larger than the threshold of the current time-slot. In other words, the subsequent time-slots lose the ability to time-order the transmissions of nodes based on their *RC* values.

To address this problem, we introduce the notion of *Time Sequence Scheme* (TSS), whose basic idea is a repeated reordering of time-slots based on decreasing threshold values. Assume that a broadcast session lasts |*T* | time-slots, which are organized in *l* levels (also referred to as *epochs*), each level (which comprises different number of timeslots) is associated with a decreasing value of the *RC* threshold. The epochs are ordered in time, with the time-slots from the uppermost level occurring first, followed by the time-slots in the next level. Also, in each level the time ordering of the time-slots is from left to right. The uppermost epoch (which contains a single time-slot) is associated with the largest threshold of *RC* (which we label as *u*)[4], allowing transmission only of nodes with *RC* value of at least *u*. The second epoch is associated with the *RC* threshold of *u*-1, allowing only nodes with *RC* values of at least *u*-1. However, as the transmission in the first epoch might have revealed new covered nodes with the *RC* value larger than *u,* the second epoch contains two time-slots: the first allowing transmission of nodes with *RC* value of at least *u,* followed by a time-slot allowing transmission of nodes with *RC* value of at least *u*-1. This process continues until the last epoch, associated with *RC* threshold of 1 and containing *u* time-slots, allows ordered transmission of nodes with *RC* values at least *u* down to nodes with *RC* values at least 1. The following diagram depicts the thresholds of the time-slots at all levels:

(*u*) ------------------------- {uppermost level *u*, threshold = *u*}
(*u*) (*u*-1) -------------------- {level *u*-1, threshold = *u*-1}
⋮                                      ⋮
(*u*) (*u*-1) … (1)  ------------{level 1, threshold = 1}

To unambiguously label each time-slot, instead of using the threshold value only as in the above diagram, we use three values: (*upper, middle, lower*). The *upper* is simply the maximum value of the threshold, *u*. The *lower* is the number of the level, and the *middle* is the actual threshold value of the time-slot.

*Algorithm 1* is used to generate the structure (*time-sequence* (*TS*)) of the |*T* | time-slots of a broadcast session, with the resulting diagram shown in Fig. 1. Upon network deployment, *Algorithm 1* is run locally by each node.[5] The *TS* properties are further formally defined in Section IV.B.

**(u, u, u)** ------------------------------------- {level *u*:  *l* = *u*}
(u, u, u-1) **(u, u-1, u-1)** --------------------- {level *u*-1:  *l* = *u*-1}
⋮                                                        ⋮
(u, u, 1) (u, u-1, 1) … **(u, 1, 1)**  ----------- {level 1:  *l*=1}

Fig. 1: The output of Algorithm 1. Time-slots in bold are called *edge slots*.

As an illustrative example, consider the case of *u* = 4 as shown in *Example 1*. In the top *TS* level (level 4) only nodes with RC ≥ 4 are allowed to transmit. In the next *TS* level (level 3), first nodes with *RC* ≥ 4 and then nodes with *RC* ≥ 3 will be allowed to transmit. In level 2, first nodes with *RC* ≥ 4, then nodes with *RC* ≥ 3, and finally nodes with *RC* ≥ 2 will transmit. In the last level, first nodes with *RC* ≥ 4, then nodes with *RC* ≥ 3, then nodes with *RC* ≥ 2, and finally nodes with *RC* ≥ 1 (all nodes with at least one uncovered neighbor) will be allowed to transmit.

**(4,4,4)** ----------------------------------------------{TS level 4: *l*=4}
(4,4,3) **(4,3,3)** -------------------------------------{TS level 3: *l*=3}
(4,4,2) (4,3,2) **(4,2,2)** --------------------------{TS level 2: *l*=2}
(4,4,1) (4,3,1) (4,2,1) **(4,1,1)** -----------------{TS level 1: *l*=1}

Example 1: The output of *Algorithm 1* for *u* = 4

The parameter *u*, is fixed and set up administratively and network-wide at the time of network deployment. The value *u* should be judiciously chosen; a too small value of *u* does not allow to separate in time the transmissions of nodes with

---

[3] This is often the case with the Greedy algorithm in various other applications as well.

[4] We discuss the choice of *u* later in the paper.

[5] Practically, *Algorithm 1* should be run periodically.

---

**Algorithm 1**: Constructing the vector set $T$

---

**Input:** $u$
**Output:** ordered collection of vectors $T = \{T_u, T_{u-1}, \ldots, T_1\}$
**Algorithm:**
1:  $T \leftarrow \varnothing$
2:  $upper \leftarrow u$
3:  $middle \leftarrow u$
4:  $lower \leftarrow u$
5:  $T_1 \leftarrow (\,upper, middle, lower\,)$
6:  $T \leftarrow T_1$
7:  **while** $middle > 1$ **do**
8:    **if** $middle == lower$
9:      $lower \leftarrow lower - 1$
10:     $middle \leftarrow upper$
11:   **else**
12:   **if** $middle > lower$
13:     $middle \leftarrow middle - 1$
14:   $T_{next} \leftarrow (\,upper, middle, lower\,)$
15:   $T \leftarrow T \cup T_{next}$

---

different values of $RC$, thus losing the ability to assign larger priority to nodes with larger $RC$ values, while a too large value of $u$ results in many empty time-slots, thus leading to an unnecessarily long broadcast session. Section V.A.2 addresses the choice of appropriate parameter $u$ in more details, discussing its effect on the scheme's performance. Next, we consider the process of node scheduling and $RC$ computation, given the $TS$ structure.

### A. Node Scheduling and Residual Coverage Computation

Given the $TS$ structure described above, each node should locally schedule its time of transmission, so that, overall, nodes with higher $RC$ transmit earlier than nodes with lower $RC$. Thus, the $TS$ serves as a common reference for all nodes.

#### 1) Local Node Scheduling
The source node transmits the message at the beginning of the first time-slot in the $TS$. As the broadcast propagates throughout the network, any node $j$ upon receiving the broadcast message for the first time determines the current $TS$ level and the current time-slot within the $TS$. This is achieved by subtracting the initial transmission timestamp (piggybacked by the broadcast message along with the broadcast message's ID) from the node's current local time, and by dividing the difference by the duration of a time-slot. Each time-slot has prefixed length and temporal format consisting of a *Preamble* part followed by a *Broadcast Field* part. The *Broadcast Field* is fixed to be the maximum duration needed to transmit the broadcast message. The *Preamble* is used to transmit control information between adjacent nodes, and its duration is negligible compared to the total length of the time-slot. Knowing the generic TS structure, as per Fig. 1, and the number of time-slots that have passed since the beginning of the $TS$, a node is able to determine the parameters of the current time-slot $T_{ct} = (upper_{ct}, middle_{ct}, lower_{ct})$. Next, node $j$ determines its residual coverage, $RC_j$, as described below.

Node $j$ then runs *Algorithm 2* to schedule its transmission for a future time-slot. Given $T_{ct}$ and $RC_j$, *Algorithm 2* schedules node $j$ to a transmission timeslot, $T_b$, later in the

---

**Algorithm 2**: Node $i$ self-scheduling

---

**Input:** $RC_j$ ; $T_{ct} = (u_{ct}, m_{ct}, l_{ct})$ // *vector associated with the current time-slot*

**Output:** transmission time-slot $T_b \leftrightarrow t_b$
**Algorithm:**
1:  $rc \leftarrow RC_j$
2:  $upper \leftarrow u_{ct}$
3:  $middle \leftarrow m_{ct}$
4:  $lower \leftarrow l_{ct}$
/* if $RC_j$ is larger than the current value of *middle*, $i$ transmits in the next time-slot */
5:  **if** $rc > middle$
6:    $T_b \leftarrow T_{ct+1}$
7:  **else**
/* if $RC_j$ is large enough for the level $T_b$ is at the current level depending on the value of $RC_j$ */
8:  **if** $rc \leq middle$ **and** $rc \geq lower$
9:    **if** $(u_{ct}, m_{ct}, l_{ct})$ **is** $edge\_slot$
10:     **if** $lower > 1$
11:       $T_b \leftarrow (\,upper, rc, lower - 1\,)$
12:     **else**
13:       $T_b \leftarrow (\,upper, rc, 1\,)$
14:   **else**
15:     $T_b \leftarrow (\,upper, rc, lower\,)$
/* if $RC_j$ is even less than the value of *lower*, $T_b$ is in a later level of the TS triangle; the level depends on $RC_j$ */
16: **else**
17: **if** $rc < lower$ **and** $rc \geq 1$
18:   $T_b \leftarrow (\,upper, rc, rc\,)$

---

broadcast session. $T_b$ could be the next timeslot immediately after $T_{ct}$ provided $RC_j$ is large enough (i.e., $RC_j > middle_{ct}$). Otherwise, *Algorithm 2* attempts to schedule node $j$ for a time-slot at the current level[6], if $RC_j \geq lower_{ct}$. Otherwise, node $j$ is scheduled to transmit at a later, lower level. In general, the larger $RC_j$, the earlier is the level and the earlier is the scheduled transmission time-slot $T_b$ within that level. If $RC_j = 0$ the node is not scheduled for transmission at all.

It is important to note that the value of a node's $RC$ can change[7] between the time at which it had scheduled itself for transmission and the beginning of its scheduled-for-transmission time-slot, $T_b$, thus possibly rendering the node inadmissible in its scheduled time-slot. To avoid transmission in an incorrect time-slot, a node re-computes its $RC$ value during the *Preamble* time of its scheduled-for-transmission timeslot and checks if it still can transmit in $T_b$. If so, it transmits the message during the *Broadcast Field* time. Else, it reschedules itself by employing *Algorithm 2* again.

#### 2) Computation of Residual Coverage
The $RC$ value of a node is needed prior to the node scheduling or rescheduling itself for transmission. This is done by a locally executed protocol – a version of a "Neighbor Discovery" protocol. Specifically, to compute its $RC$ value, node $i$ transmits a *Coverage Request* packet, *CReq*, to all its 1-

---

[6] If the current time-slot is an edge slot (see Fig. 1), *Algorithm 2* attempts the next level of the time-sequence.

[7] Due to transmissions of other nodes or due to mobility.

hop neighbors. The *CReq* packet contains *i*'s ID and the ID of the broadcast message[8]. As the *CReq* packet contains the ID of the queried message, a node can easily verify whether it has or has not received the message before (i.e., whether the code is *covered*, or *uncovered*, respectively). Upon receiving the *CReq* packet, each one of *i*'s uncovered neighbors replies with a *Coverage Reply* packet, *CRep*, which contains the neighbor's ID, *i*'s ID and the broadcast message ID. Node *i* then counts the number of such replies during a time approximately equal to the duration of the *Preamble*. Notice that compared to a typical data broadcast message these control messages are rather short.[9]

### B. Formal Properties of the Time Sequence

Here we derive more formally the properties of the *TS* structure based on *Algorithm 1*. We also illustrate how, along with the scheduling *Algorithm 2*, the TSS scheme emulates the behavior of the centralized greedy algorithm. We start by discussing the complexity of *Algorithm 1*.

Let *T* be an ordered collection of vectors $\{T_x, T_{x-1}, ..., T_1\}$, where $T_i = (u_i, m_i, l_i)$, where $u_i$, $m_i$, and $l_i \in \mathrm{N}^+$, and where the parameters *u, m,* and *l* are the *upper, middle,* and *lower* values. Let *T* be the output of *Algorithm 1* given input *u*.

*Proposition 1:* The time complexity of *Algorithm 1* is $O(u^2)$; and the length of the generated time sequence is $|T| = \dfrac{u(u+1)}{2}$.

*Proof:* The output, *T*, of *Algorithm 1* can be arranged in an isosceles triangle with sides *u*. The triangle consists of *u* levels, where the last level (level 1) comprises *u* vectors. The number of vectors at the $i^{th}$ level equals $1 + u - i$. The total number of vectors is, then: $\sum_{i=1}^{u} i = u(u+1)/2$. At each iteration of *Algorithm 1* there is exactly one vector generated; hence there are $O(u^2)$ iterations and $|T| = u(u+1)/2$. □

As discussed in Section V.A.2, the value of *u* in practice is rather low and does not depend on the network density. Hence, *Algorithm 1* is not computationally expensive and could be run on resource-constrained nodes.

Next, we formally define the structure of the *TS*.

Let $\mathrm{A} = (S, \mathrm{I})$ be a set system where the ground set *S* is the set of all network nodes. Each node *j* computes its value $p_j = RC(j)$. More specifically, $S = \{n_1^{RC_1}, n_2^{RC_2}, ..., n_N^{RC_N}\}$, where $|S| = N$. As time goes by from a time-slot to the next, the values $RC(j)$ may change. Let I be the collection of subsets of *S*, such that $\mathrm{I} = \{S_1, S_2, ..., S_x\}$ and $S_i = \{s_j^{p_j} : p_j \geq m_j \geq l_j \geq 1\}$ given vector $T_j = (u_j, m_j, l_j)$ in *T* (note that $|\mathrm{I}| = |T|$).

Every *x* consecutive time-slots (naturally ordered in time) are mapped one-to-one to the vectors in the ordered collection *T* above. That is, each time-slot, $t_i$, is uniquely associated with a vector in *T*: $t_1 \leftrightarrow T_x = (u, u, u)$, $t_2 \leftrightarrow T_{x-1} = (u, u, u-1)$, …, $t_x \leftrightarrow T_1 = (u, 1, 1)$, where $x = 0.5u(u+1)$ from *Proposition 1*.

Define a binary order relation on *S*, such that

$s_j^{p_j} \leq s_i^{p_i} \Leftrightarrow p_j \leq p_i$.

*Definition 5:* An element $s_j^{p_j}$ is called *admissible in $T_i$*, iff $s_j^{p_j} \in S_i$.

A network node is admissible in $t_i$ if it is admissible in $T_{x+1-i}$. This association "wraps around"; i.e., in general for $i \geq 1$, $t_i \leftrightarrow T_{mi+1-i}$ where $m = \lceil i/x \rceil$. The *TS* is the ordered collection of the time-slots, together with their corresponding vectors $T_i$.

*(Revisited) Definition 1:* A *broadcast session* consists of all the events, starting from the transmission of the message by the source node and ending when the broadcast algorithm terminates.

Also, for now, we will assume that the broadcast session finishes in less than or equal to $|T|$ time-slots.

*Broadcast Rule:* At every time-slot, $t_i$, a node considers transmitting only if it has not transmitted earlier in this session and it is admissible in $T_i$ in the *Preamble* of $t_i$.

It is easy to verify that *Algorithm 2* complies with the *Broadcast Rule*. For instance, if $m_{ct}$ of the current *TS* time-slot is lower than the node's *RC* value, *Algorithm 2* schedules the node's transmission for the next immediate time-slot. Otherwise, a further future time-slot is assigned to the node. A node is never scheduled to transmit if its $RC = 0$ (i.e., it is not admissible in any time-slot).

Define a minimal *T* admissible element in $T_k$, denoted by $min(T_k)$, to be $s_j^{p_j}$ such that $s_j^{p_j} \leq s_i^{p_i}$ for all $s_i^{p_i}$ admissible in $T_k$. Consider the ordered collection of vectors $T^\lambda$ at the level $\lambda$ of *T* ($\lambda \in \{1, ..., u\}$), and the smallest element of all minimal admissible elements at this level, denoted $inf(T^\lambda)$. Now, let *L* be the sequence $(inf(T^u), inf(T^{u-1}), ..., inf(T^1))$.

*Proposition 2:* The sequence *L* is decreasing.

*Proof outline:* The result follows trivially from the definition of I, the ordering of *T*, the order relation defined on *S*, and the definition of $inf(T^\lambda)$. □

That is, given the TS structure and scheduling algorithm, covered nodes with smaller *RC* values would not be able to transmit (i.e., be admissible) at the beginning of the time sequence *TS*, but will potentially be admissible in later timeslots (i.e., at lower levels). And reversely, only nodes with larger *RC* values would be admissible and be able to transmit in earlier timeslots (i.e., at higher levels) of the *TS*.

Let *M* be the sequence $(min(T_k), min(T_{k-1}), ..., min(T_1))$, where $T_i$'s, $i \in \{1, ..., k\}$, are in $T^\lambda$ for some $\lambda$.

*Proposition 3:* The sequence *M* is decreasing.

*Proof outline:* Similarly to *Proposition 2*, the proof follows trivially from the definition of I, the ordering of *T*, the order relation defined on *S*, and the definition of $min(T_k)$. □

Then, within given level of the *TS*, nodes with smaller *RC* values would again be admissible only in later timeslots at this level, and nodes with larger *RC* values would be admissible earlier in the given level.

In summary, the structure of the *TS* (implemented via *Algorithm 1*) in conjunction with the *Broadcast Rule* (implemented via *Algorithm 2*) has admissibility property allowing for repetitive assignment of larger transmission priority to nodes with larger *RC* values as compared to nodes with smaller *RC* values. Hence, by the virtue of the admissibility property of the *Broadcast Rule*, the greedy "oracle" scheme is approximately emulated.

---

[8] Each node remembers the IDs of the broadcast messages that it has received. Note that the length of time that a node needs to retain the list of IDs of received broadcast messages is limited to the duration of a broadcast only, which typically would be several seconds in a practical scenario.

[9] This simple protocol can be further improved in a variety of ways, but this is beyond the scope of this paper and is not considered here further.

### C. Time-Sequence Based Broadcast Schemes

For clarity, let's summarize the basic *TS*-based broadcasting scheme. All nodes run *Algorithm 1* to construct the *TS*. When a node transmits the message, all of its previously uncovered neighbors mark themselves as covered, compute their *RC*, and run *Algorithm 2* to schedule their transmission time-slots. Just before its scheduled time to transmit (during the *Preamble* of the scheduled-for-transmission time-slot), a node re-computes and updates its *RC*. If the node's current value of *RC* has decreased (but $RC > 0$), the node determines its new time-slot assignment by re-running *Algorithm 2*. If, at any time, the *RC* value of a node equals 0, the node will never be scheduled for transmission in this broadcast session.

We refer to this basic scheme as the *Naïve Time Sequence Scheme* (*NTSS*)[10]. In what follows, we present a variant of the NTSS, utilizing 1-Hop neighborhood topology knowledge.

#### 1) The 1-Hop Time-Sequence Scheme (TSS)

Suppose node $i$ is about to transmit[11] in the current time-slot $t_k$. Then node $i$ checks during the *Preamble* of the time-slot $t_k$ whether any of its 1-hop neighbors are scheduled to transmit within $t_k$ as well. This check does not necessitate any additional transmissions, as the node can determine whether a particular neighbor is scheduled to transmit in $t_k$, if it receives the *CReq* message from that neighbor during the *Preamble* time. If more than one neighboring node is scheduled for $t_k$, the node with the largest *RC* is selected to transmit in $t_k$. The rest of the neighboring nodes reschedule themselves to transmit in the next time-slot. (To accommodate the TSS operation, the *CReq* and the *CRep* control packets should also include the respective sender's *RC* value, in addition to the nodes' IDs.)

One could also considered a 2-Hop time-sequence based scheme, where node $i$ checks during the *Preamble* of the time-slot $t_k$ whether any of its 2-hop neighbors are scheduled to transmit within $t_k$ as well. However, our study demonstrated that such a scheme does not perform significantly better than its 1-Hop counterpart. Indeed, it performs worse in dynamic topologies due to the larger number of control packets that are transmitted; the same observation holds for network with packet loss. Even in the case of static, collision-free networks the benefits of a 2-hop *TS*-scheme are also not significant.

### D. Sample Execution of NTSS

Consider the network of nodes shown on Fig. 3. For simplicity of presentation, only the NTSS variant of the *TS* scheme is discussed, which suffices to demonstrate the main mechanisms of the *TS* scheme. Suppose $u = 4$. Then, *Algorithm 1* constructs the *TS* as shown in *Example 1*. The sample run of NTSS is shown in Fig. 2, with the resulting network coverage depicted in Fig. 3.

Note that the *TS* structure allows *Algorithm 2* to give priority to nodes with higher *RC* that are scheduled *later* during the broadcast session. For instance, because of its

| Scheduled Nodes | RC/Scheduled Timeslot |
|---|---|
| *a* | 2 / (4,2,2) |
| *b* | 3 / (4,3,3) |
| *c* | 1 / (4,1,1) |
| *d* | 1 / (4,1,1) |

**Step 1**

| Scheduled Nodes | RC/Scheduled Timeslot |
|---|---|
| *a* | 2 / (4,2,2) |
| *c* | 0 / (4,1,1) |
| *d* | 1 / (4,1,1) |
| *f* | 1 / (4,1,1) |
| *g* | 0 / - |
| *h* | 0 / - |

**Step 2**

| Scheduled Nodes | RC/Scheduled Timeslot |
|---|---|
| *c* | 0 / (4,1,1) |
| *d* | 1 / (4,1,1) |
| *f* | 0 / (4,1,1) |
| *g* | 0 / - |
| *h* | 0 / - |
| *e* | 0 / - |
| *k* | 2 / (4,2,1) |

**Step 3**

**Step 4:** All nodes are covered after timeslot (4,1,1). The time-sequence is exhausted at (4,1,1), where as well no nodes have RC > 0, and the algorithm terminates.

Fig. 2: In the first timeslot, the source node *S* transmits the message. Nodes *a*, *b*, *c*, and *d* receive it, mark themselves as covered, compute their RC, and schedule themselves to broadcast At **Step 1**, according to *Algorithm 2*, since node *b* has 3 uncovered neighbors it is scheduled for timeslot (4,3,3). Node *a* is scheduled similarly for (4,2,2), *c*, and *d* for (4,1,1). At **Step 2**, during the third timeslot (4,3,3) node *b* transmits the message. Nodes *g*, *h*, and *f* become newly covered and are added to the scheduled nodes list: *f* is scheduled for (4,1,1); *g* and *h* have RC = 0 and are not scheduled. Node *b* is removed from the list. At **Step 3**, in the sixth timeslot (4,2,2), node *a* check its *RC*. Since its *RC* remains the same, node *a* transmits the message. Nodes *e* and *k* become newly covered. Node k has two uncovered neighbors: *i* and *j*. It is scheduled for timeslot (4,2,1). Finally, at **Step 4**, during the preamble of the ninth timeslot node *k* has not changed its RC and transmits. All nodes are covered at this point. Hence, *c*, *d*, and *f* do not transmit.

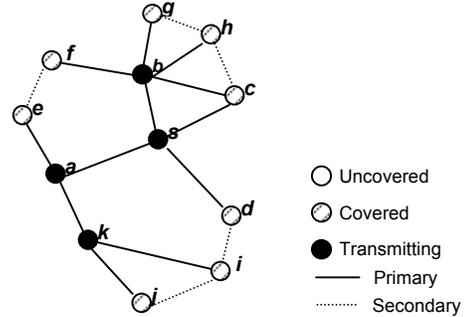

Fig. 3: A network topology example, where the TS-based scheme picks {*s*,*b*,*a*,*k*} and form a MCDS of the particular network graph. All nodes are covered after the broadcast session completes.

larger *RC* value, node *k* will transmit before nodes *c*, *d*, or *f*, albeit they both received the broadcast message and were scheduled earlier. Ultimately, this eliminates the transmissions of nodes *c*, *d*, and *f*. Fig. 3 shows the network state after the broadcast session is completed. Edges between transmitting nodes, and between transmitting and covered nodes are labeled as primary; edges between covered nodes as secondary.

### E. Correctness of the TS-based Schemes

In the following proof of correctness, the network topology is assumed to be static[12] during the broadcast session and the MAC layer to be perfect. The network graph is assumed to be connected[13]. The performance evaluation of the TSS scheme in the next section, though, demonstrates that even in the presence of high mobility and severe packet loss, the scheme achieves full or almost full network coverage.

---

[10] As the name indicates, the NTSS possess some deficiency, which will be cured by the other variant of the scheme: TSS.

[11] I.e., that the current value of RC(*i*) matches the time-slot $t_j$.

[12] For an arbitrary mobility model even flooding cannot ensure full network coverage; similarly, this is true if the package loss probability is positive.

*Proposition 4:* The *TS*-based schemes terminate in finite amount of time and guarantee full coverage of the network.

*Proof outline*: Per *Algorithm 2*, for all *TS* schemes, a node is not scheduled to transmit unless its *RC* is strictly greater than zero. Whenever a node *n* transmits, all of it neighbors receive the broadcast message and are marked as covered. Hence, the *RC* value of node *n* decreases to zero, and node *n* is not admissible in any future time-slot. Therefore, a node does not transmit more than once during the execution of the algorithm. Since the number of nodes in the network is finite, the algorithm terminates in a finite number of time-slots.

Now, suppose that after a *TS*-based algorithm's termination there is at least one node, *D*, that is not covered. Since the network graph is connected, there exist at least one path from the source node, *S*, to the destination node *D*. Then, because *D* has not received the message, there are at least two neighboring nodes *X* and *Y* along this path, such that *X* has received the message and *Y* has not (note: *X* might be *S* and *Y* might be *D*). Therefore, *X*'s *RC* was greater than zero at algorithm's termination. This contradicts the *TS* schemes' termination condition. (Whenever a node is covered it computes its residual coverage. As long as node's residual coverage is greater than zero it is always scheduled to transmit according to *Algorithm 2*.) Thus, by contradiction, the *TS* based schemes cover all the network nodes. □

## V. PERFORMANCE EVALUATION

We investigate the performance, defined by a number of metrics, of various broadcast algorithms in four distinct network topology models. For a static network topology, we consider both the cases of a perfect MAC-layer (no packet loss due to collisions) and the case of positive packet loss probability. Also, we consider two types of mobile topologies: one with *independent* mobility patterns of the nodes; and another with *correlated* (group) mobility patterns.

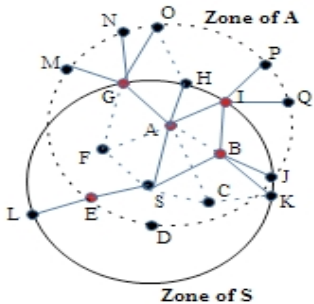

Fig. 4: Zone radius *R*=2. The Bordercast source node *S* broadcast the message to all its neighbors *A*, … ,*F*. However, only *A*, *B*, and *E* are selected to rebroadcast since they cover all of *S* border nodes: *G*, …, *L*. Similarly, considering next *A* as the rebroadcast node, only *G* and *I* are selected among *A*'s neighbors to cover all of *A*'s uncovered border nodes *M*, …,*Q*. Note that even though *J*, *K* and *D* are also border nodes of *A* they are already covered.

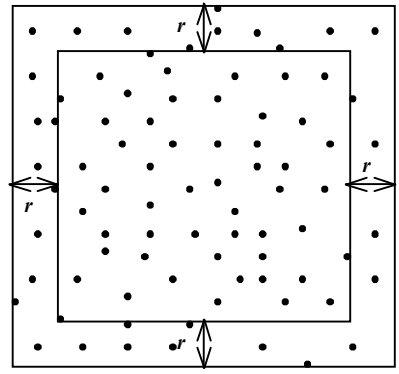

Fig. 5: Double boundary simulation area; *R* denotes the transmission range of a node.

We compared the performance of the TS-based schemes against the most efficient schemes found in the technical literature to the best of our knowledge, at the time this paper has been written. In particular, we simulated the *position-aware Responsibility Based Scheme* (*RBS*), suggested by Khabbazian and Bhargava [25]. In [25] authors show that RBS outperforms a few well known broadcast algorithms, the *Edge Forwarding* [13] algorithm, for example. Another broadcast protocol simulated here, the *Bordercast,* is the route discovery mechanism in the *Zone Routing Protocol* (*ZRP*) ([26]). *Bordercast* relies only on local topological information to select the nodes to forward the broadcast message. Per ZRP (see Fig. 4), a zone of node *A* in the network includes all nodes that are within *R* hops from *A*. Border nodes are those nodes in the zone whose minimum hop distance from *A* is exactly *R*. According to the *Bordercast* algorithm, the goal is to cover most efficiently only all of the border nodes in its zone. Fig. 4 illustrates the basic workings of *Bordercast*. Within the context of ZRP, it is guaranteed that the entire network would still be covered [26] even though a fraction of the nodes in the zone will be pruned and will not transmit the message.

To include an algorithm that constructs a *backbone* structure prior to the broadcast session, we selected the *Funke's* algorithm [21], which was shown to obtain one of the best approximation ratios to the size of the MCDS, outperforming leading algorithms, such as the *Wan's* algorithm [20]. Finally, for comparison, we also simulated the Liu's algorithm [11] – a *node forwarding* algorithm that relies on 1-Hop positional information. Both RBS and Funke's algorithms are among the most efficient broadcast schemes in the literature to date.

In all the experiments, the simulation area is a 200[m] by 200[m] square; the inner square area is of dimensions (200 - *r*)[m] by (200 - *r*)[m], where *r* [m] is the transmission radius of all nodes and was set to *r* = 25[m] (see Fig. 5). The number of nodes in the network was varied to investigate the schemes' performance at different node densities. All the broadcast algorithms were implemented in a JAVA discrete event simulator.

### A. Static Network Topology with Lossless MAC

#### 1) Transmission Complexity

The number of transmission during a broadcast session (i.e., "transmission complexity"), a crucial metric for an efficient broadcast algorithm, is investigated in Figures 6 a) and b), in the case of the lossless MAC layer. We were interested in understanding how our distributed algorithms compare with the centralized greedy approximation of the MCDS.

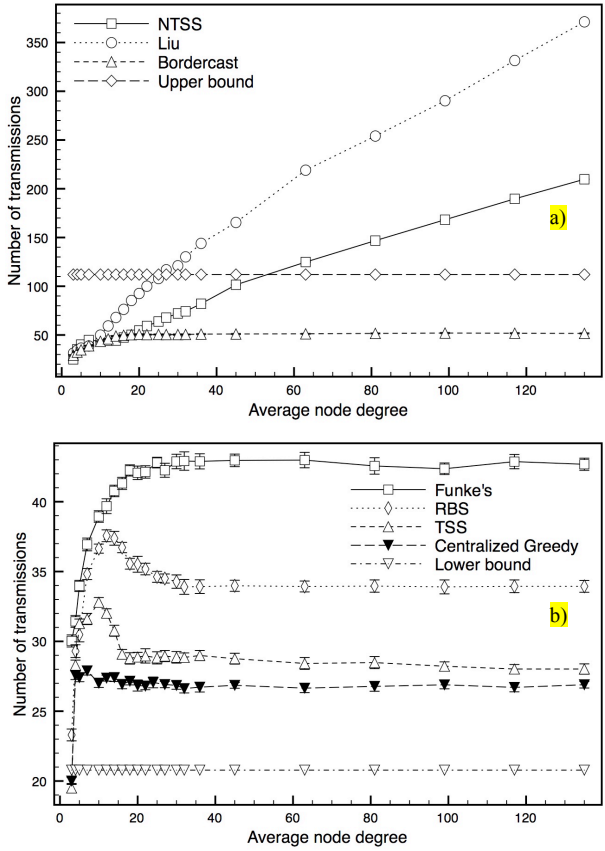

Fig. 6: The number of transmissions in a static collision free network (full coverage).

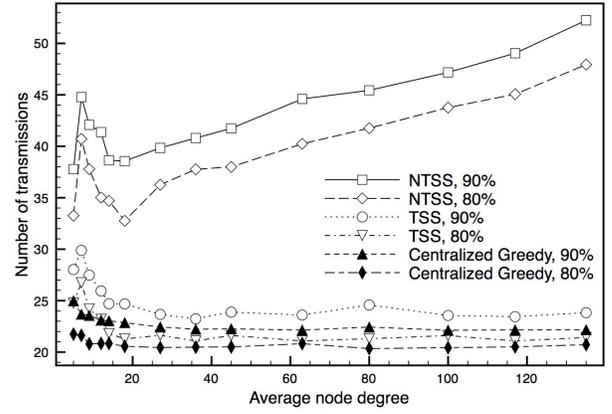

Fig. 7: The number of rebroadcasts to achieve fractional coverage of 80% and 90% in a static collision free network.

As an additional comparison of the transmission complexity, we calculated the number of transmissions using the Linear Hexagon Coverage technique, which provides a close approximation to the minimum number of transmissions needed to cover the entire network area, assuming sufficient node density. The upper bound of transmission complexity should be interpreted as the maximal number of transmissions that would be required by any broadcast algorithm that avoids duplicate coverage of nodes and, hence, is *density-independent*. More detailed discussion of these bounds can be found in Section VI.

Both, the NTSS and the Liu's algorithms are not *density-independent* (surpass the upper bound). The NTSS algorithm does not perform well, because often a node is covered multiple times by transmissions of its 1-hop neighbors which occur in the same time-slot, thus resulting in redundant transmissions. This is especially common in time-slots towards the end of a broadcast session. The number of such transmissions during a simulation run is proportional to the number of nodes in the network and, hence, the transmission complexity continues to increase with the node density.

So far, all simulation results were in the context of 100% coverage of the network nodes. However, in some practical scenarios such full coverage is not essential. Fig. 7 demonstrates the performance of the TS schemes, with the Greedy algorithm as benchmark, respectively assuming 80% and 90% coverage suffices. As expected, the needed number of transmissions decreases for all schemes. This decrease is most significant for the NTSS scheme, due the previously described phenomenon of inefficient transmissions towards the end of the time sequence. When only fractional coverage is needed, most of such inefficient transmissions do not occur because the time sequence is truncated before its last timeslots.

### 2) *Delay*

The delay – the number of time-slots needed to complete a broadcast session – is presented in Fig. 8. The delay performance of the three TS-based schemes was obtained with $u$ set to the smallest value, so that the number of broadcast transmissions is still close to optimal. Thus, it could be considered as the worst case scenario. These smallest values of $u$ were determined via simulations (Figures 9 (a), (b) and (c)); it is noticeable that beyond average node degree of 15, the $u$ values for TSS remain essentially constant with respect to the network density. Thus the upper value could be fixed prior to network deployment, resulting in the delay and the number of broadcast messages close to the algorithm's optimum behavior. The results for TSS hint that a good practice would be to set $u$ to the average node degree for networks of node degrees up to 20. For networks of larger

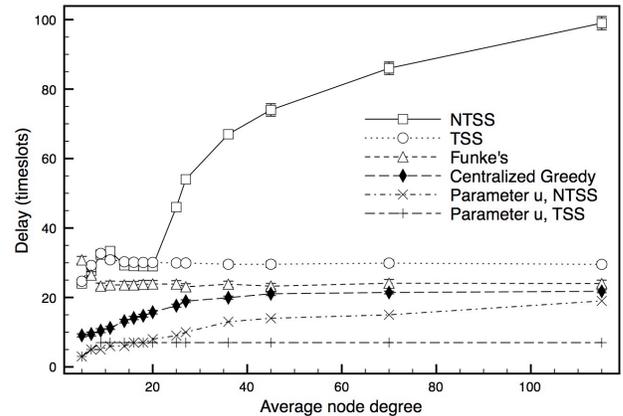

Fig. 8: The broadcast delay and the corresponding values of the parameter $u$; for the static, collision free, full coverage case

node degrees, $u$ is largely density-independent and can be simply fixed at 20.

In general, there is a tradeoff between the two performance metrics: the number of transmissions during a broadcast session and the duration of a broadcast session. Intuitively, it would be beneficial to set $u$ to a large value for finer-grain resolution of priorities among the nodes with different $RC$ values and, thus, for a better approximation to the greedy "oracle" scheme. However, large values of $u$ have the disadvantage of increasing the delay, since the TS grows longer (quadratic in $u$, from *Proposition 1*); also large number of initial time-slots go by empty, as no nodes are admissible then.

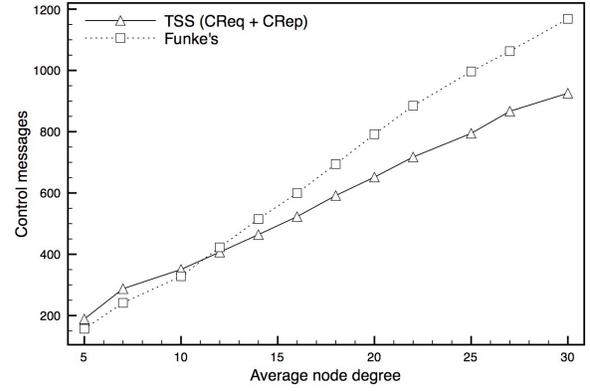

Fig. 10: Control messages per TSS and Funke's algorithms; both schemes' control messages are comparable in size and short.

### 3) Control Messages

Given the optimal values of the parameter $u$, we study the performance of TSS in terms of control messages (CMs) overhead. We also compare the number of CMs generated by TSS with that generated by the Funke's algorithm. Both TSS and Funke's algorithms' CMs are comparable in size. In TSS, all CMs are generated during the *Preamble* of a time-slot; the length of the *Preamble* is negligible compared to the *Broadcast Field* since each CM is very short.

### B. Dynamic Network Topology

For all of the mobility experiments, each data point was obtained after a simulation warm-up time of 1000 [$s$] in which the mobility model could converge to a stationary node distribution. The number of nodes $N = 200$.

### 1) Gauss-Markov Node Mobility

TABLE 1
THE PARAMETERS OF THE GAUSS-MARKOV MOBILITY MODEL

| | |
|---|---|
| *Velocity and position update interval* | 0.2 [s] |
| *Velocity standard deviation* | 0.75 [m/s] |
| *Velocity mean* | Varied |
| *Alpha* | 0.75 |

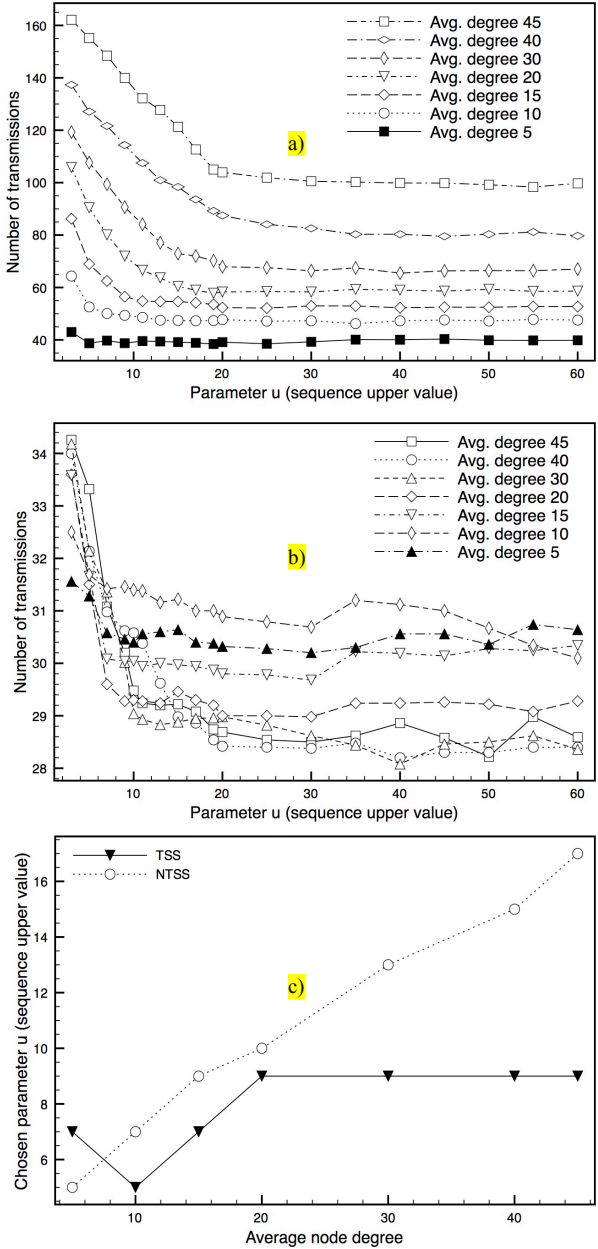

Fig. 9: Determining the optimal values of the parameter $u$, so that the trade-off between number of rebroadcasts and delay is optimized. Figure a) NTSS, b) TSS, and c) the chosen optimal values used throughout the paper for the schemes NTSS and TSS.

In the *Gauss-Markov Mobility Model* (GMMM) [33, 34], time is split into time intervals (independent of the TS-based schemes time-slots). At the beginning of the $k^{th}$ time interval, nodes' velocity is updated based on their velocity in the $(k-1)^{th}$ time interval and according to the following rule: $v_k = \alpha v_{k-1} + (1-\alpha)\bar{v} + \sqrt{(1-\alpha^2)}v_{x_{k-1}}$. Here, $v_{k-1}$ is the velocity (speed and direction) of a node in the previous time interval, $v_{x_{k-1}}$ is a Gaussian random variable, $\bar{v}$ is the mean value of the velocity, and $\alpha$ is a parameter that determines the degree of which the current velocity depends on the previous velocity (the amount of "memory"). As $\alpha$ approaches 1, nodes' motion becomes more constant; as $\alpha$ approaches 0 nodes' motion becomes more random. At the end of each interval, the position of a node in the network area is updated according to its velocity during this interval. Table 1 summarizes the values of the parameters.

Fig. 11 depicts the number of transmissions generated by the three investigated broadcast algorithms. As the average speed of the nodes increases, the number of transmissions saturates for all algorithms. This is an artifact of the finite network area − beyond a certain speed (about 20-30 [m/s] in this case), the locations of the nodes in the next time interval become random with respect to the locations in the previous time interval. This is equivalent to formation of a completely new random placement of nodes in each time interval, thus increasing the speed further does not alter the performance of the algorithms anymore.

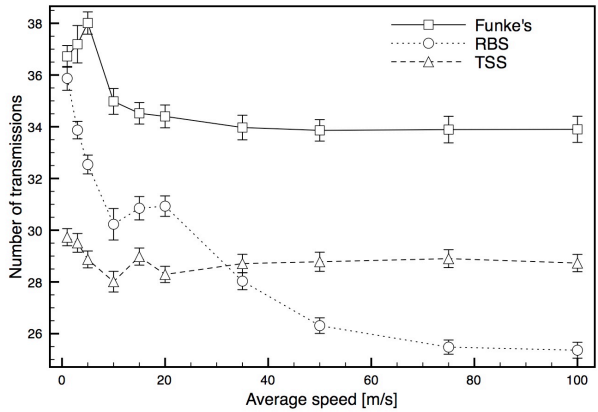

Fig. 11: The number of transmission for the GMMM case; $N = 200$ nodes

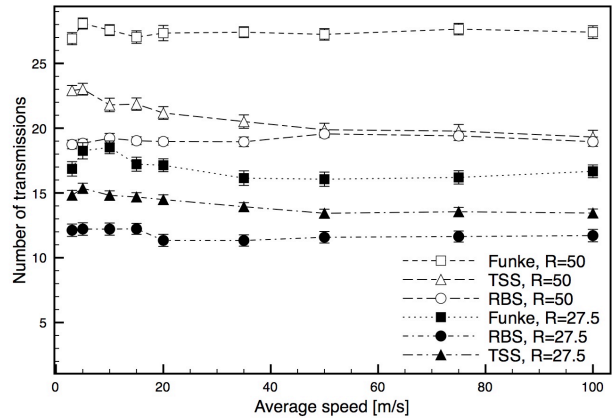

Fig. 13: Number of transmission for the RPGM case; $N = 200$

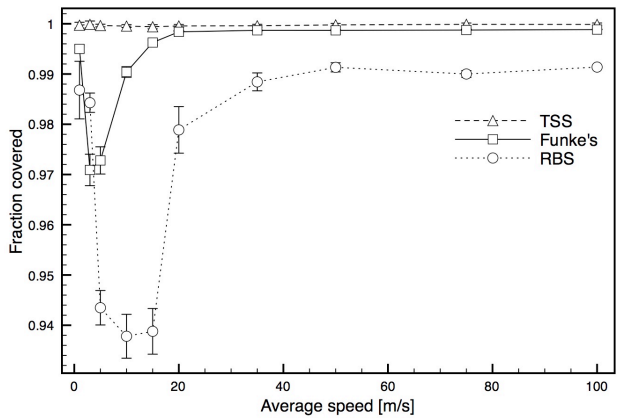

Fig. 12: The fraction of nodes covered for the GMMM case; $N = 200$

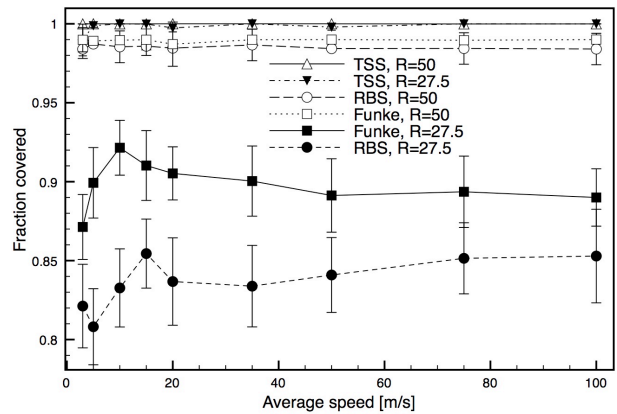

Fig. 14: Fraction of covered nodes for the RPGM case; $N = 200$

However, the number of transmissions prior to this saturation point depends on the specific mechanisms of the broadcast algorithm in question. For instance, the decrease in the number of transmissions prior to saturation as mobility increases for the *TS*-based schemes results from the scheduling policy of these algorithms: the larger their residual coverage is, the sooner they transmit. As mobility increases, there is higher chance that a node with lower *RC* (scheduled for later timeslots) would move into an area containing larger concentration of uncovered nodes, thus effectively increasing its *RC* and, overall, reducing the total number of transmissions. Furthermore, the fraction of nodes covered increases as well, as shown in Fig. 12.

As opposed to the static case, where a broadcast session is bound to cover all the network nodes as long as the network is connected, with mobility there is a chance that some nodes will not be covered. Fig. 12 demonstrates the coverage

(fraction of the network nodes reached during a broadcast session). The performance of all algorithms saturates beyond some mobility threshold. In the mobility region of 10-20 [m/s] and for smaller densities (200 nodes in the network), all algorithms are affected by graph partitions during the broadcast session.

The notable drop in performance of the RBS scheme in Fig.12 for a range of low speeds is due to the specific RBS mechanisms: node *A* transmits if among its neighbors there is an uncovered node *B*, such that the distance between *A* and *B* is smaller than the distance between *B* and any other of the *covered* neighbors of *A*. For very low velocity, RBS performs well and covers almost the entire network; however, as the velocity increases, node *A*'s estimation of the distance between *B* and *A*'s covered neighbors becomes less accurate. Though, as the velocity of nodes increases further, RBS benefits from the fact that *A* could move into areas where only a small number of nodes are covered.

On average, RBS achieves slightly lower network coverage and, consequently, slightly lower number of transmissions than the *TS*-based schemes, while the latter obtain almost complete coverage in this scenario.

### 2) Group Mobility

Since independent individual node movement may be unrealistic for certain scenarios (e.g., hikers, tourist groups, military platoon, UAVs, etc.), we have also simulated a group mobility model, which is based on the *Reference Point Group Mobility* (*RPGM*) model described in [35]. Here, the implemented version of RPGM follows the *Nomadic Community Mobility* model described in [34]. In this model the network nodes form groups that move together from one reference point (*RP*) in the simulation area to another. The *RP*s have a radius of influence, and the group of nodes is free to move randomly within this radius. The simulation parameters (such as the group sizes and the number of groups in the network) used here were borrowed from [34] and [36] for the Nomadic and the RPGM models, respectively. The number of nodes was set to 200. Table 2 lists the parameters' values.

TABLE 2
THE PARAMETERS OF THE RPGM MOBILITY MODEL

| Pause time for mobility around | 0.3 [s] |
|---|---|
| Nodes velocity update frequency | 0.5 [s] |
| Maximum reference point pause | 4 [s] |
| RP radius | Varied |
| Average group size | 20% of network size |
| Nodes mean velocities | Varied |

Figures 13 and 14 depict the number of transmissions and the network coverage. The *RP* radius (labeled as *R* in Fig. 13, 14), is varied to investigate the effects of node clustering about the reference point. As can be observed, the smaller is the *RP* radius, the more frequent are the network partitions and, consequently, the performance of all algorithms degrades.

As in the GMMM case, so in the RPGM case too, the number of transmissions and the network coverage saturate as the network mobility increases. However, in the RPGM case there is a minimal decrease in the number of transmissions with respect to the increase in mobility; e.g., for 27.5[m] the decrease in the number of transmission is largely due to the highly clustered and partitioned network topology.

On average, RBS again generates fewer transmissions, but also covers a smaller fraction of the network, as compared to the TSS scheme in Figures 13 and 14. Overall, in dynamic network scenarios, although TSS relies only on local topology, it tends to outperform backbone-based or position dependent algorithms such as Funke's and RBS, respectively.

### C. Performance with Packet Loss (Lossy MAC)

In this section we investigate the effect of packet loss (e.g., due to MAC-layer collision) on the performance of TSS in a static network. The performance of TSS with packet loss is quite robust in terms of delay and coverage (Figs. 16 and 17). Although the number of transmissions needed to cover (almost) the entire network increases as the probability of packet loss increases (Fig. 15), however, even in severe packet loss cases (e.g. 20% packet loss), the number of TSS transmissions is still comparable to the collision free performance of backbone-based schemes. For instance, in a <u>collision-free</u> network, Funke's backbone algorithm requires on the average 43 transmissions, while in a network with 20% packet loss, TSS requires on average 44-45 transmissions. Similarly, TSS with 20% packet loss outperforms Liu's and the *Bordercast* algorithms running in a <u>collision-free</u> network. The RBS scheme operating in a collision-free network requires approximately 34-37 transmissions to cover the network; in contrast, this is the number of transmissions

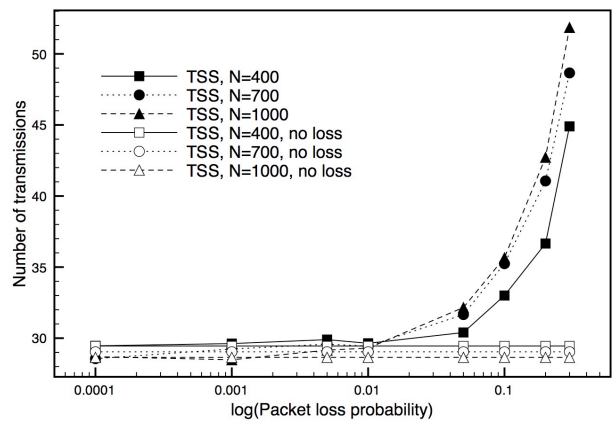

Fig. 15: Number of transmissions vs. packet loss for the TSS scheme

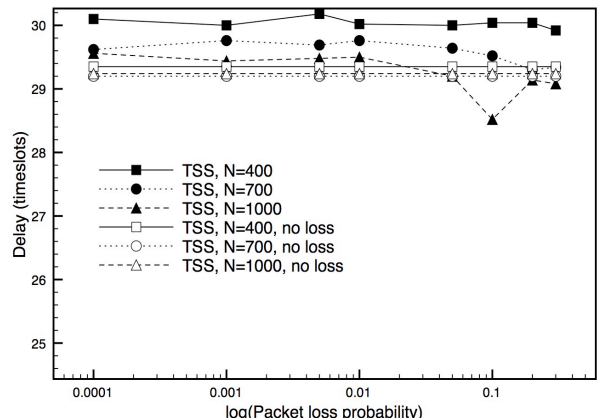

Fig. 16: Delay vs. packet loss for the TSS scheme

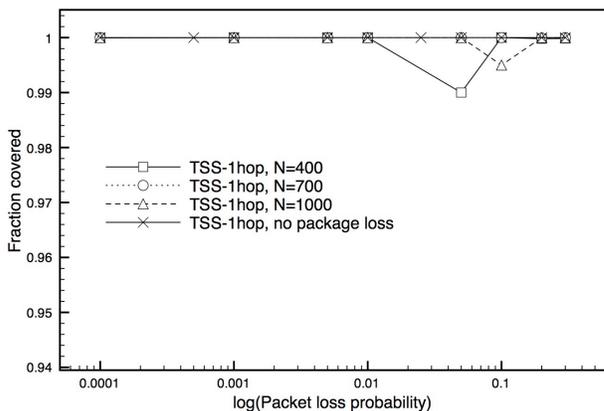

Fig. 17: Fraction of covered nodes vs. packet loss for the TSS scheme

needed by the TS schemes to cover the network with up to 15% loss probability.

Simultaneously, the delay of the TS-schemes is not affected much by the packet loss, since in this case some of the slots that were not utilized in the collision-free network, are now used to compensate for the packet loss. The results hold for different node densities in the network ($N = 400, 700, 1000$).

### VI. SUMMARY AND DISCUSSION

In this paper, we introduced a novel scheme for broadcasting in wireless networks that mimics in performance of the centralized greedy algorithm. This is accomplished through distributive prioritization of transmissions based on nodes' residual coverage (*RC*) and the particularly designed Time Sequence (*TS*) to schedule the nodes' transmissions. The basic NTSS scheme was improved to eliminate a major source of inefficiency − multiple coverage of nodes by more than one transmission − which resulted in the TSS scheme. We proved the schemes' correctness (i.e., guaranteeing full coverage in finite time in the case of static and collision-free network).

Through simulations and based on two metrics − the transmission complexity and the delay − we compared the performance of our schemes with other leading broadcasting schemes, initially in static networks. The TSS scheme outperforms all other schemes with respect to the number of message transmissions, without requiring additional equipment, such as GPS. Furthermore, this performance is achieved with bounded latency, and is independent of network density. Finally, the performance of the TSS scheme is robust even in network characterized by severe packet loss.

Next, we considered two types of mobility scenarios: individual node and group mobility. We showed experimentally that TSS outperforms the other schemes with respect to the network coverage.

Our study allows examination of the basic tradeoffs in the design of the broadcasting protocols, such as the tradeoff between transmission complexity and delay. It is worth noting that the TSS schemes facilitate control of this tradeoff through a single parameter only − the *u* parameter.

Even though theoretical bounds have been suggested before in the literature, in the current study we derived simple general upper and lower bounds on the number of retransmissions of any broadcast algorithm operating over a finite-area network. The bounds are suitable for broadcast algorithms benchmark, sparing the brute force computation of |MCDS| ([41], [42]), and are independent of knowledge of the size of the Maximal Independent Set. In the process, we quantified the notion of "reasonably performing broadcast algorithm," thus allowing the quick identification of inefficient algorithms surpassing certain bound on *reasonable* number of transmissions.

## VII. RELATED WORK

The problem of efficient broadcasting has been extensively studied in the technical literature. The initial simple concept of flooding evolved to more sophisticated schemes through building optimal network subgraphs. All through, the main algorithmic challenge has been to reduce the number of transmitted messages needed to reach all the network nodes.

Among the major shortcomings of pure flooding are the large transmission complexity and the notorious broadcast storm [1]. The *Scalable Broadcast Algorithm* [2] alleviates somewhat this problem utilizing 1-Hop neighbor information.

A different approach is taken by probabilistic broadcast protocols that associate some (re)transmission probability to each node receiving the broadcast message. Schemes exploring such mechanisms were suggested in [3]-[9].

The interest in probabilistic broadcasting schemes is due to their inherent low transmission overhead, low processing complexity, and high tolerance of frequent and rapid topological changes. Balancing these benefits, though, is the disadvantage of inability to guarantee full network coverage.

In contrast, deterministic broadcast algorithms innately guarantee full network coverage (assuming ideal MAC layer). In the deterministic scheme of Multipoint Relaying proposed in [10], the set of retransmitting neighbor nodes is reduced from the set of all neighbors to the minimum subset of neighbors that cover the same area as that covered by the original set. This approach is an example of the "minimum forward-node set" strategy, and works such as [11]-[14] provide approximate solutions to this NP-hard problem. To avoid the transmission of the list of forwarding nodes along with the broadcast message, the technique of self-pruning [15]-[16] has been proposed.

The forward-node set and, consequently, the self-pruning problems can essentially be viewed as the task of solving the NP-hard "Minimum Connected Dominating Set" (MCDS) problem [13]. Several studies ([18]-[22]) have attempted to tackle the problem by constructing a communication backbone prior to the broadcast initiation (see [46] for comparison of such approaches). These schemes can sometimes dramatically reduce the number of transmissions. Nevertheless, as shown in [23], they do not tolerate well frequent network topological changes. For volatile communication environments, an approach to dynamically construct CDS is a better alternative. Works such as [24], [25], and the present study offer potential solutions. However, further research is required to study the scalability of these approaches.

Bounds on the size of |MCDS| in unit disk graphs (UDG) have been studied extensively in the literature. Typically these bounds are given via ratio between the cardinalities of the Maximal Independent Set (MIS) and MCDS. Let $\mu = |MCDS|$ and $\chi = |MIS|$. Wu et al. demonstrate that $\chi \leq 4\mu + 1$ [41]. Further improved band is derived in [42], where $\chi \leq 3.43\mu + 4.82$. Finally, a tight (non-improvable) bound is derived by Vahdatpour et al.: $\chi \leq 3\mu + 3$, [43]. These bounds on |MCDS| are helpful provided |MIS| can be found efficiently. The authors of [44] suggest an optimal distributed algorithm that can be applied for UDG (asymptotically optimal with running time O(log*n)). A Polynomial-time Approximation Scheme (PTAS) for finding a Maximum Weight Independent Sets is provided in [45].

A number of studies have attended to the theoretical complexity bounds of broadcast and related information dissemination mechanisms for topology (not area) bounded networks. For instance, [37, 38] provide such complexity bounds on gossiping, broadcast, and rumor spreading in such networks. Also, influential works by Peleg et al. in the spirit of [39] demonstrate important complexity lower bounds on broadcast in radius-2 radio networks: the broadcast procedure requires $\Omega(\log^2 n)$ transmissions.

## APPENDIX: BOUNDS ON WIRELESS BROADCAST

In this section upper and lower asymptotic bounds on the number of retransmissions in a 2D wireless ad hoc network are demonstrated based on geometrical considerations (given the system model described in Section II). The bounds are general and hold for any broadcast algorithm over UDG.

Typically the bounds on the size of MCDS for UDG are presented in terms of an approximation ratio between the cardinality of the Maximal Independent Set (MIS) and the

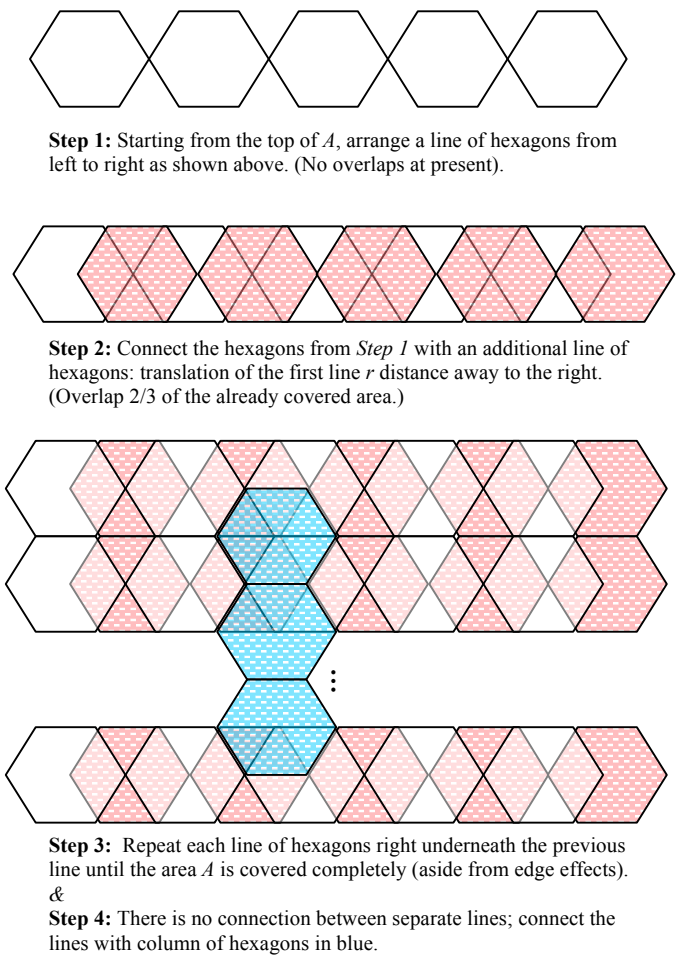

**Step 1:** Starting from the top of $A$, arrange a line of hexagons from left to right as shown above. (No overlaps at present.)

**Step 2:** Connect the hexagons from *Step 1* with an additional line of hexagons: translation of the first line $r$ distance away to the right. (Overlap 2/3 of the already covered area.)

**Step 3:** Repeat each line of hexagons right underneath the previous line until the area $A$ is covered completely (aside from edge effects). & **Step 4:** There is no connection between separate lines; connect the lines with column of hexagons in blue.

Fig. 18: Construction of the Linear Hexagon Packing

cardinality of the MCDS (Section VIII provides more details and references). In contrast, the bounds shown below are simple, useful for the purposes of algorithms benchmarks, and are characterized by closed form expressions depending only on the size of the area enclosing the network and nodes' transmission radius, $r$. This is useful for high density networks since computing the exact size of MCDS is NP-hard problem ([32]), which of course is intractable for large number of nodes, and the running of algorithms (be they efficient) finding constant factor approximations either to MIS or MCDS for every graph instance may neither be feasible nor required.

Let the minimum number of nodes transmitting a broadcast message (equivalently number of transmissions) be $\beta_{\min}$, the number of nodes in the network be $N = |S|$, the area enclosing the network be $A$, its size be a constant $|A|$, the set of all points in $A$ covered by transmission of node $i$ be $\lozenge_i$, and $\mu \equiv |\text{MCDS}|$. As noted earlier, $\beta_{\min} = \mu$.

Let the area $A$ be rectangular. Given graph topology $\Lambda(S)$, the goal is to pick[14] a set of nodes $X \subseteq S$ positioned so that $i$) $\bigcup_{i \in X} \lozenge_i$ covers $A$ entirely; $ii$) $\bigcap_{i \in X} \lozenge_i$ is minimized (i.e. the overlap between transmissions is minimal); and $iii$) for any two nodes $i$ and $j$ in $X$, $i$ and $j$ can *communicate*: $\exists$ a path of nodes entirely in $X$ connecting $i$ and $j$. Let $\Phi_{\Lambda(S)}$ be the conjunction of $i$), $ii$), and $iii$) given topology $\Lambda(S)$.

### A. Lower Bound on the Number of Transmissions

As $N \to \infty$, $\Phi_{\Lambda(S)}$ yields a recast of the MCDS problem with solution the set $X$ in the context of a *bounded* area $A$ and UDGs. Namely, $|X| = \beta_{\min}$.

Now, let $\lozenge_I = D_i$ where $D_i$ are identical disks with radius $r$ (i.e. all nodes in the network have the same transmission range). Under the constraint of $\Phi_{\Lambda(S)}$, let $|X_D| = |X| = \beta_{\min}$. It is easy to check that utilizing circular disks $D_i$ to satisfy $i$) of $\Phi_{\Lambda(S)}$ would introduce substantial overlaps between the disks covering area $A$. Instead, suppose each of the $N$ disks $D_i$ is approximated with a hexagon $H_i$ with side $r$ and centered at $i$. That is, under the constraint of $\Phi_{\Lambda(S)}$, let $\lozenge_I = H_i$ and let $|X_H| = |X|$ in this case. The plane is readily tessellated by hexagons thus avoiding the overlaps needed to satisfy $i$). In essence, $\beta_{\min} = |X_D| > |X_H|$ as long as a construction is provided so that $ii$) and $iii$) are also satisfied, given $\lozenge_I = H_i$.

Consider the Linear Hexagon Packing (LHP) construction as described in Fig. 18. A node from $S$ is picked to be in $X$ if it is at a centroid of a hexagon in the LHP. The LHP results from arranging hexagons in multiple lines. In each line, every two adjacent hexagons share only one common point. A node needs to be picked at every such point so that all nodes covered in each line of hexagons can communicate, per $iii$) of $\Phi_{\Lambda(S)}$, with least possible resulting overlap between hexagons, per $ii$) of $\Phi_{\Lambda(S)}$. It is easy to check that now only 2/3 of each hexagon's area is covered twice. Yet, at this point nodes covered in line $x$ can communicate only with other nodes covered in $x$. Placing an additional hexagon for every two adjacent lines is sufficient so that all $N$ nodes communicate (and $ii$), $iii$) of $\Phi_{\Lambda(S)}$ are satisfied). Notice that $i$) of $\Phi_{\Lambda(S)}$ is still satisfied by the LHP construction at this point.

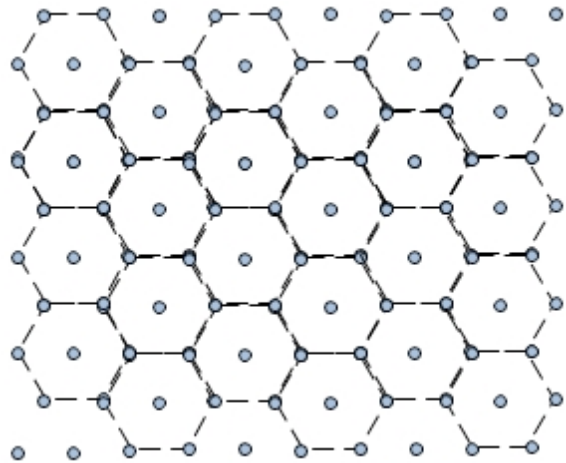

Fig. 19: Honeycomb-based topology: nodes are placed at the vertices and centers of tessellating hexagons of side $r\sqrt{3}$.

Suppose the area $A$ is a square of side $d$ and let $q = \sqrt{|A|}/r$ (assuming without loss of generality $d/r$ is an integer). Asymptotically then, employing LHP, accounting for the 2/3 overlap and the additional connecting hexagons, one obtains

$$\beta_{\min} > k = |X_H| = (q^2 + q - \sqrt{3})/\sqrt{3} \qquad (1)$$

### B. Upper Bound on the Number of Transmissions

We start with a useful definition emphasizing a property of reasonable broadcast algorithms.

*Definition 5*: A *reasonable* broadcast algorithm generates at most $K(q)$ transmissions.

Notice that $K(q)$ is independent on the network density; that is, a reasonable algorithm scales as the number of nodes in $A$ increase. For instance, flooding is *not* a reasonable broadcast algorithm by *Definition 5* since the number of transmissions is proportional to the number of nodes in the network. Notice that a trivial upper bound on the number of retransmissions of all *reasonable and unreasonable* broadcast algorithms (such as flooding) could be $N$: the number of nodes in the network. Here we'd like to obtain $K(q)$ instead, which is an upper bound on the number of transmissions of *reasonable* broadcast algorithms. Next, we discuss the Worst Case Topology Construction.

Suppose we can position the nodes in the network at will. Then, a topology can be constructed that maximizes the minimum needed number of nodes, $N$, whose transmission disks would cover the entire network area and result in a connected graph. That is, given rectangular area $A$ and a number of nodes $N$ (not necessarily infinite), one seeks graph with topology, $\Lambda(S)$ such that

$$\beta_{\min} = \max |X|, \qquad (2)$$

where as before $X \subseteq S$ is the set of nodes picked for transmission while $\Phi_{\Lambda(S)}$ is satisfied; $\diamond_i = D_i$ here. Notice that the following properties hold given $X$ and $\Lambda(S)$ satisfy (2): first, $|X|=N-1$ (the last covered, $N^{\text{th}}$, node is not picked to transmit); second, it is guaranteed that the addition of new nodes in the network area would not require additional transmissions since every new node would be positioned in area already covered by one of the $N-1$ nodes in $X$; and third, if $N$ is only dependent on $A$'s dimensions and $r$, then $N-1$ is equal to $K(q)$ by *Definition 5*.

In the previous section, to construct an approximation of the lower bound on the number of transmissions we employed linear arrangement of hexagons. Here, to construct $\Lambda(S)$ a different hexagons' property would be utilized.

In 1940, L. Fejes Tóth [40] proved that the densest packing of circles (or any collection of other shapes given each shape encloses the same surface area) in the plane is obtained by the honeycomb hexagonal lattice.

Suppose the plane is tessellated by hexagons. Then, the honeycomb circle packing requires that each packed circle is centered at the centroid of a tessellating hexagon. The largest number of packed circles results. Figure 19 demonstrates the union of 7 such circle packings (the center of each circle in the union of packings is represented by a node in Fig. 19); only one of the underlying 7 honeycomb lattices (dashed line) is shown in Fig. 19 for clarity.

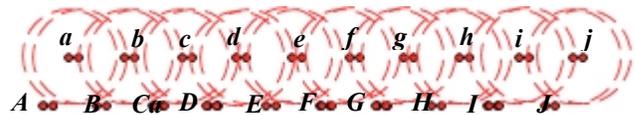

Fig. 20: Removal of any node from pairs $a$ through $j$ would render the communication between nodes in the broadcast row impossible. Also note that the nodes from pairs $A$ through $J$ are not covered by pairs $a$ though $j$ tx's.

Again, here we assume, without loss of generality, that the rectangular area is a square with side $d$ and $|A|=d^2$; the radius of a circle is equal to the transmission radius $r$, but the side of a tessellating hexagon this time is equal to $r\sqrt{3}$.

Suppose each node in Fig. 19 is replaced by a pair of nodes separated by distance $\varepsilon$, where $\varepsilon \to 0$.

Consider the transmissions' row in Fig. 20. The removal of any node from the row would disconnect the graph. Note that at this point, nodes in row $x$ of the hexagonal topology are not connected to nodes in rows other than $x$ (i.e., the rows form connected components in the graph). To connect the rows into single connected component, additional nodes need to be placed. As above, we would like to maximize the minimum number of connecting nodes needed.

Figure 21 demonstrates the final graph topology $\Lambda(S)$ including the additional connecting pairs of nodes (in green). Note that eliminating any node's transmission (except of the last covered node) would cause disconnections in the graph. Hence, $N$-1 nodes need to transmit so that condition *iii*) of $\Phi_{\Lambda(S)}$ is satisfied; thus indeed $|X| = N$-1. Given $N$-1 transmitting nodes *i*) and *ii*) of $\Phi_{\Lambda(S)}$ are also satisfied; *ii*) is satisfied since due to symmetry, one cannot select a different subset $X$ of the $N$ nodes, such that $\bigcap_{i \in X} \diamond_i$ is less.

By construction, $\Lambda(S)$ maximizes the minimum number of nodes needed whose transmission disks would cover the entire network area and result in connected graph. Initially there are $r-1$ pairs per row; adding the nodes needed to connect every two rows yields $r$ +1 pairs per row. Since the rows are distance

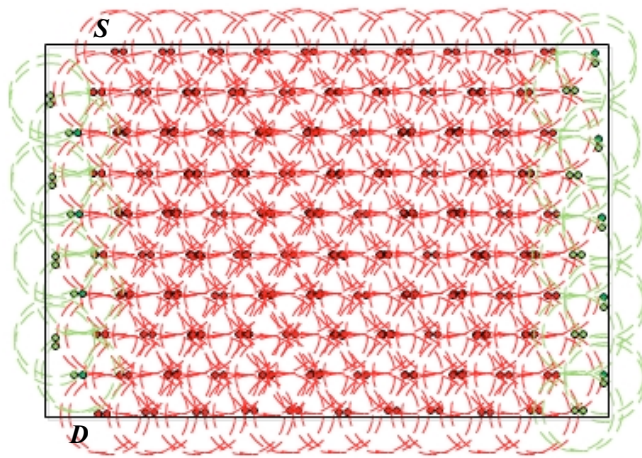

Fig. 21: Nodes in green connect broadcast rows so that all $N$ nodes communicate. Broadcast starting from node $S$ proceeds in a "zig-zag" fashion until node $D$ is reached. All $N$ nodes are required to transmit; $|MCSD|=N$. Removal of any node would cause disconnection. The network area (black rectangle) is completely covered (barring edge effects).

$r$ from one another, there are $r-1$ rows. Consequently, at most

$$\beta_{min} = |X| = N-1 = 2(q+1)(q-1) - 1 = 2(q^2-1) - 1 \qquad (3)$$

transmissions are needed for network coverage. Hence, $|X|$ is independent of the network density. By *Definition 5*, $K = |X| = \beta_{min}$, providing an upper bound on the number of needed transmissions by any *reasonable* algorithm to cover the network.

$K$ is also the lowest number of transmissions any algorithm can *guarantee* given a network area with dimensions $d$ x $d$ and transmission radius of $r$. In contrast, the lowest number of transmissions any algorithm can ever achieve as the density of the network grows to infinity is approximated by (1).